\documentclass[floatfix,
  aps,% American Physical Society
  prapplied,% Physical Review Applied
  reprint,% two-column journal style
  superscriptaddress,% APS-style affiliations
  longbibliography% PR journals prefer full titles in refs
]{revtex4-2}
\usepackage{float}
\usepackage{graphicx}% Include figure files
\usepackage{dcolumn}% Align table columns on decimal point
\usepackage{bm}% bold math
\usepackage{siunitx}
\usepackage{hyperref}% add hypertext capabilities
\usepackage{orcidlink}

\begin{document}
\raggedbottom
\preprint{APS/123-QED}
\title{Cell Reproduction in a Dark Optical Trap}% Force line breaks with \\
%Claw-like Optical Tweezer applied to cells
\author{Ariel Hertz\,\orcidlink{0009-0003-3184-8960}}
\email{arielhertzdearaujo@gmail.com}
\affiliation{Department of Physics, Pontifical Catholic University of Rio de Janeiro, Rio de Janeiro 22451-900, Brazil}
\author{Gabriel Dias\,\orcidlink{0009-0001-6935-5585}}
\affiliation{Department of Physics, Pontifical Catholic University of Rio de Janeiro, Rio de Janeiro 22451-900, Brazil}
\author{Carlos L. R. Fragoso\,\orcidlink{0000-0001-9495-3889}}
\affiliation{Department of Chemistry, Pontifical Catholic University of Rio de Janeiro, Rio de Janeiro 22451-900, Brazil}
\author{Anna De Falco\,\orcidlink{0000-0001-9549-2711}}
\affiliation{Centro Nacional de Ressonância Magnética Nuclear, Departamento de Bioquímica Médica, ICB/CCS/UFRJ, Rio de Janeiro, 21941-590, Brazil}
\author{A. Z. Khoury\,\orcidlink{0000-0002-7487-5067}}
\affiliation{Instituto de Física, Universidade Federal Fluminense, Niterói, RJ, 24210-346, Brazil}
\author{Thiago Guerreiro\,\orcidlink{0000-0001-5055-8481}}
\email{thguerreiro@gmail.com}
\affiliation{Department of Physics, Pontifical Catholic University of Rio de Janeiro, Rio de Janeiro 22451-900, Brazil}

\begin{abstract}
Optical tweezers are a versatile tool in the domain of cytology, enabling trapping and manipulation of individual cells. However, the incidence of laser light causes photodamage to biological matter, even when operating at low optical powers and over short periods of time. Here, we demonstrate stable trapping of single living \textit{Saccharomyces cerevisiae} yeast cells \textit{in vitro} using a dark optical trap operating in the repulsive regime of light–matter interactions. In contrast to standard tweezers, our dark optical trap confines cells for hours with negligible disruption to their morphology and reproduction cycle, opening up new possibilities for long duration experiments with living organisms such as the observation of cell reproduction under laser trapping. 
%propose and demonstrate an alternative optical trapping technique exploiting repulsive light–matter interactions: the Claw Dark Optical Trap. We trap single cells \textit{in vitro} with minimal optical damage, tunable for a wide range of cell shapes and sizes. We demonstrate optical trapping in the called claw dark tweezer using three Gaussian beams of equal power to trap \textit{Saccharomyces cerevisiae} yeast cells. We show that yeast cells remain confined for hours with minimal disruption to their morphology and reproduction cycle, in contrast to cells trapped in a conventional Gaussian tweezer.
\end{abstract}
\maketitle

\section{\label{sec:level1}Introduction}

With their ability to hold and displace objects using light, optical tweezers (OTs) enable the manipulation of nano- and micron-sized objects with exquisite control and precision \cite{ashkin1986observation,neuman2004optical,bustamante2021optical}. Relying on the transfer of momentum between light and matter, OTs have found a wide range of applications, including thermodynamical  studies of mesoscopic systems \cite{trepagnier2004experimental,collin2005verification}, nonlinear dynamics \cite{gieseler2014dynamic,kremer2024perturbative}, many-body physics \cite{rieser2022tunable,penny2023sympathetic}, quantum mechanics \cite{pikovski2012probing,weiss2021large,dania2025high,rademacher2026roto}, and the manipulation of biological samples \cite{Ashkin1987,Ashkin1987-2,Keloth2018,wilson2025free,Qu2011}.

The force exerted by an OT upon an object depends on the relative refractive index $m = n_{p}/n_{m}$ between the particle and the medium \cite{Volpe2023,millen2020optomechanics,gonzalez2021levitodynamics}. The most common operation regime is given by $m > 1$, in which particles are attracted to the maximum of intensity of the beam. Alongside a conservative trapping force, there is scattering and absorption of light, which immediately raises an issue: energy is absorbed by the trapped object, possibly leading to laser damage. For inanimate objects this is generally not a problem, but for living organisms such as cells and bacteria, it can lead to significant damage \cite{mohanty2005dynamics,pilat2017effects,blazquez2019optical,zhang2008optical}. As well, absorption and photon recoil are limiting factors for quantum experiments with levitated particles \cite{jain2016direct, hackermuller2004decoherence}.

Beyond the usual Gaussian tweezer, several works have explored structured light, particles and optical media to create optical trap configurations operating in the regime of negative relative polarizability, given by $m < 1$. In this case, optical force becomes repulsive rather than attractive, and by engineering the tweezers's optical mode it is possible to realize Dark Optical Traps (DOT), which minimize scattering and energy absorption by the trapped object \cite{PhysRevApplied.14.034069, PhysRevLett.131.163601,illetschek2026dark, afridi2026controlling}.

\begin{figure}[ht!]
    \centering
    \includegraphics[width=0.3\textwidth]{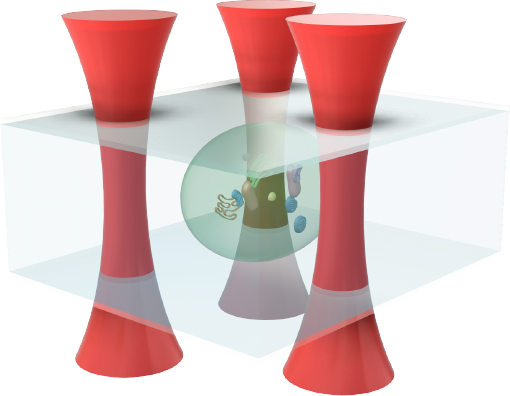}
    \caption{\textit{Saccharomyces cerevisiae} yeast cell trapped in the claw dark tweezer.}
    \label{fig:ilustracao}
\end{figure}

To overcome the photodamage caused by optical tweezers in cell manipulation, and inspired by dark optical traps operating in the repulsive regime, we implemented an alternative trapping scheme based on displaced Gaussian beams, creating a stable trapping point where incident light is negligible, as conceptually pictured in Fig.~\ref{fig:ilustracao}. This \textit{claw dark tweezer} can be shaped to match the geometry and size of the cells, providing a versatile confinement and manipulation platform.

In this work we show optical trapping in the claw dark tweezer, formed by three Gaussian beams of equal power, to trap S. cerevisiae cells with a near-infrared laser and a high numerical aperture objective.

%In this configuration the size of the trapping region and the magnitude of the exerted force are decoupled, so that high numerical aperture lenses can be used to achieve stronger confinement without significant change. -- Comentei essa parte, muito tecnica pra introducao.

%\section{Trapping}
\section{Repulsive optical forces}

To create the claw dark trap, we must reach the repulsive regime of the optical forces governed by the relative refractive index of the particle and medium. For near-IR light, the refractive index of \textit{S. cerevisiae} cells lies in the range $n_{\mathrm{cell}} = 1.36 \sim 1.39$\cite{liu2016cell,rappaz2005measurement}. Reaching $m < 1$ therefore requires a medium with a refractive index above this range. At the same time, the medium must preserve cell viability: since photodamage is assessed here through the reproduction of the trapped cells, the medium itself must not inhibit cell division.

The chosen medium is a solution of water, optimal YPG (yeast extract-peptone-glucose \ref{app:solution_composition}) and Iodixanol. Water and YPG supply the nutrients required for the survival and reproduction of the yeast, but have refractive indices below the target value ($n_{\mathrm{water}} = 1.333$ and $n_{\mathrm{YPG2x}} = 1.343$, measured in a refractometer at $24^\circ$C), so Iodixanol was added to raise the index of the solution.

Iodixanol, commercially available as Optiprep (60\% Iodixanol, 40\% water), is a density gradient medium with a refractive index of 1.429, already well established for cell isolation \textit{in vitro} \cite{Boothe2017}. Being non-ionic, metabolically inert and iso-osmotic, it raises the refractive index of the medium without imposing osmotic stress on the cells, which makes it particularly suitable here. Its influence on cell viability was evaluated separately (Appendix \ref{app:solution_composition}), and a proportion of Iodixanol, YPG and water was then chosen so that the effects of the trapping technique could be analyzed independently of the effects of the medium.

Nine different solutions were prepared, with refractive indices ranging from 1.338 to 1.426; their compositions are listed in Appendix B and numbered from 1 to 9. In the repulsive regime, the larger the difference between the refractive indices of the cell and of the medium, the stronger the force exerted by the laser, but the solution must also retain enough YPG to sustain reproduction. Solutions 1 to 7 produced clear repulsive interactions, although the reproduction capacity of the yeast decreases as the YPG content is reduced, eventually ceasing altogether. The repulsive character of the interaction is shown directly in Movie S1 \cite{supplemental}, in which a single Gaussian beam is brought towards a cell and pushes it away from the focus. Solution 7 ($n = 1.401$) was adopted for all trapping and viability reproduction experiments, as it offers the best compromise between the strength of the optical forces and the nutrient content.

In solution 8 ($n = 1.397$) the repulsive behavior was no longer unambiguous: the relative refractive index is too close to the threshold between the repulsive and attractive regimes, and small inhomogeneities of the cell membrane are enough to switch the interaction from one regime to the other. Solution 9 contains no Iodixanol and lies in the positive-polarizability region, so that the cells are attracted to one of the Gaussian beams; it was used to perform the comparison experiments with a single Gaussian tweezer.

\begin{figure}[ht!]
    \centering
    \includegraphics[width=\columnwidth]{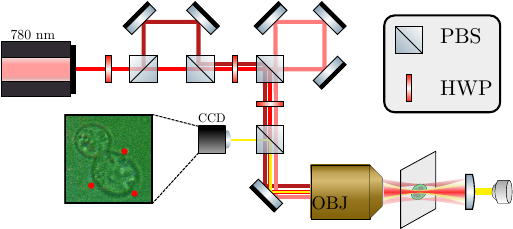}
    \caption{Experimental setup of the claw dark tweezer. The red path represents the laser, which is split into three independent beams by the polarizing beam splitters (PBS); the half-wave plates (HWP) balance the power among them. The three beams are then focused by the objective (OBJ), each focus being displaced independently from the center. The yellow path represents the imaging system, illuminated by a white-light source and collected by a CCD camera.}
    \label{fig:setup}
\end{figure}

We further observed that the refractive index of \textit{S. cerevisiae} depends on the maturity of the cell due to the difference in its composition \cite{bianco2023label}. Repulsive interactions are easier to observe with mature cells, whereas younger cells occasionally display attractive behavior even in the solutions with the highest Iodixanol content.

\section{Experimental setup}

The experimental setup is illustrated in Fig.~\ref{fig:setup}. A continuous-wave (CW) laser of wavelength $\lambda = \SI{780}{\nano\meter}$ (Toptica DL-pro) seeds a tapered amplifier (Toptica BoosTa). The amplified beam is split into three independent beams by half-wave plates (HWP) and polarizing beam splitters (PBS), which also allow the power of each beam to be balanced individually. Each beam carries approximatelly 100 mW of power, and are focused by an oil-immersion objective lens of NA $= 1.3$ (Olympus UPlanFLN 100x, adjustable NA $= 0.6-1.3$). The sample plane is imaged onto a CCD camera, with a calibrated scale of \SI{0.053}{\micro\meter} per pixel, so that the cells can be monitored and manipulated in real time. The solution containing the yeast is placed between two microscope slides positioned at the focal plane of the objective.

Yeast cells morphology varies dynamically, differing from one another in shape, density and size, with diameters typically ranging from 3 to \SI{8}{\micro\meter}. The Maltese cross interference pattern produced by the high-NA focus at the microscope slides \cite{Novotny2012} is used to align the three beams and to displace each focus individually from the center, so that the resulting trap can be matched to the geometry of the cell and made tight enough to confine it. For this reason, the three foci are not arranged in an equilateral triangle: their positions are adjusted independently for each cell, and the resulting asymmetry of the trap is reflected in the confinement measurements.

\begin{figure*}[!t]
    \centering
    \includegraphics{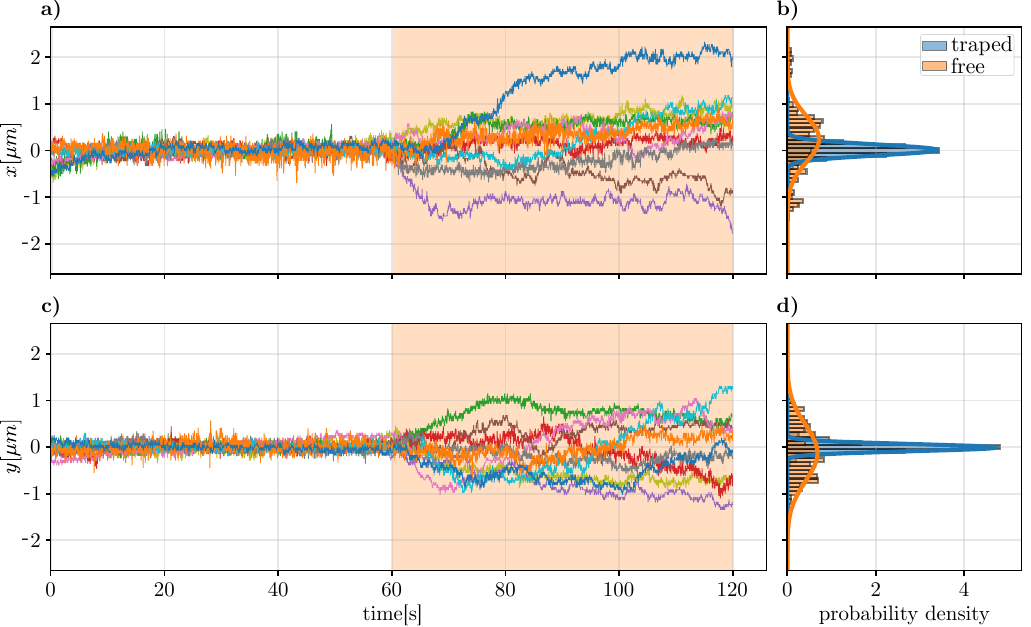}
    \caption{Confinement of single yeast cells in the claw dark tweezer. (a, c) Transverse coordinates $x$ and $y$ of ten individual cells as a function of time, each curve corresponding to one cell and referred to the mean position of that cell while trapped. During the first 60 s the tweezer is on and the motion remains confined to a region around the center of the trap; in the shaded region the laser is switched off and the cells spread in random directions, driven by Brownian motion and by their own active motion. (b, d) Corresponding distributions of the transverse position, accumulated over all cells, for the trapped (blue) and free (orange) intervals, together with Gaussian fits. Narrowing of both distributions under trapping quantifies the confinement seen in (a) and (c).}
    \label{fig:confinement}
\end{figure*}

\section{Cell Confinement}

After alignment of the trapping beams, confinement is demonstrated through a differential measurement: the cell is held in the claw dark tweezer for one minute, after which the laser is switched off and the cell is left free for one further minute. A recording of one such realization is provided as Movie S2 \cite{supplemental}. Cell positions are extracted from the recorded videos using the object-detection model described in Appendix \ref{app:yolo}.

The dynamics of the cell are shown in Figures~\ref{fig:confinement}(a) and (c). During the first 60 s the motion is confined to a region around the center of the trap; once the laser is switched off, the cells spread in random directions, driven by Brownian motion and by the active locomotion in response to chemical and thermal stimuli.

The corresponding position distributions are shown in Figs.~\ref{fig:confinement}(b) and (d). The variances along $x$ and $y$ differ slightly while the cells are trapped, as expected from the asymmetry of the three-beam configuration, from the shape of the cell and from its active motion. When the trap is switched off, both distributions broaden considerably, in agreement with the traces in Figs.~\ref{fig:confinement}(a) and(c).

The divergence of the beams is responsible for confining particles along the optical axis $z$. Motion along the optical axis is perceived in the CCD camera as the object moving in and out of focus. We can therefore infer $z$ confinement by observing a trapped cell over extended periods of time such as many hours (see Fig \ref{fig:trapped_claw_reproduction} below), for which the image remains focused over all of the experiment.
%However, because of the proximity of the microscope slides it was not possible to verify whether this effect alone is sufficient to confine the cell in the axial direction.

\section{Reproduction and photodamage}

\begin{figure}[ht!]
    \centering
    \includegraphics{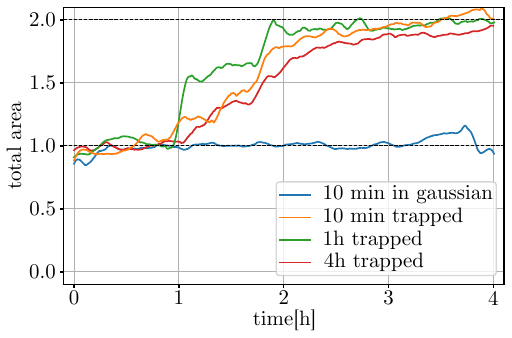}
    \caption{Cell reproduction under laser trapping. Each curve represents the mean of 5 yeast cells growing, being trapped for the specified time. We have three curves showing the results of cells being trapped in the claw dark tweezer, each of them for different times of trapping, and one curve to compare with the well known regular Gaussian tweezer in the positive polarizability regime.}
    \label{fig:growing_graph}
\end{figure}

\textit{S. cerevisiae} is one of the most extensively studied eukaryotic organisms, and its budding cycle is well characterized \cite{herskowitz1988life,botstein2011yeast}. Because its reproduction dynamics are predictable, any deviation from the expected budding behavior can be attributed to the stress imposed on the cell, which makes reproduction a sensitive indicator of the damage caused by an optical tweezer. Photodamage in a single Gaussian trap has been modeled and evaluated in Ref.~\cite{Keloth2018} as a function of the energy absorbed by the cell, which in turn depends on trapping time and laser power. The effects of this photodamage can be assessed by analyzing the visible deformation of the cell and its reproduction cycle, using the time required for a bud to form and the duration of the budding event as parameters.

In Ref.~\cite{pilat2017effects} the authors propose that the total area (the visible area of the trapped yeast cell plus the area of the cell born from it during budding) is a better indicator of the cellular stress caused by photodamage. Using this parameter, it has been shown that at low powers and short trapping times the reproduction rate is not visibly affected, but the Gaussian tweezer nevertheless deforms the trapped cell, which visibly shrinks, and the cells generated from it are considerably smaller than expected and occasionally deformed. At higher powers and longer trapping times, the trap can arrest cellular activity and reproduction altogether \cite{pilat2017effects,mohanty2005dynamics}.

In the claw dark tweezer, all beams are displaced from the center where the cell is held, and they interact appreciably with it only at the edges of the membrane, when the cell attempts to move away from the central position. The cell is therefore expected to experience only the Gaussian tails of the beams, whose contribution can be neglected, since the total absorbed energy in this configuration is minimal.

To observe the effect of the tweezer, individual cells were trapped for a fixed time and their growth was then followed, for a total of 4 hours per experiment including both the trapping time and the subsequent passive observation. For the claw dark tweezer the trapping times were 10 minutes, 1 hour and 4 hours; in the last case the entire observation window is carried out with the trap on. These were compared with cells trapped for 10 minutes in a single Gaussian beam in the positive-polarizability regime. The claw dark tweezer experiments were performed with each of the three beams at \SI{100}{\milli\watt}, giving a total power of \SI{300}{\milli\watt}, in solution 7 (Appendix \ref{app:solution_composition}); the Gaussian tweezer experiments used a single \SI{30}{\milli\watt} beam in solution 9, which provides the attractive regime.

The two schemes therefore operate in different media, and the effect of the medium alone on cell viability was characterized separately (Appendix \ref{app:cell_viability}). Since solution 9 contains no Optiprep, its YPG concentration is higher than that of solution 7, so that the cells used in the Gaussian measurements are in fact in a more favorable environment for reproduction than those used in the claw dark tweezer measurements. The reproduction of cells in both solutions in the absence of laser illumination is reported in Appendix~\ref{app:cell_viability} and provides the laser-free reference for the curves discussed below.

\begin{figure}[t!]
    \centering
    \includegraphics[width=0.48\textwidth]{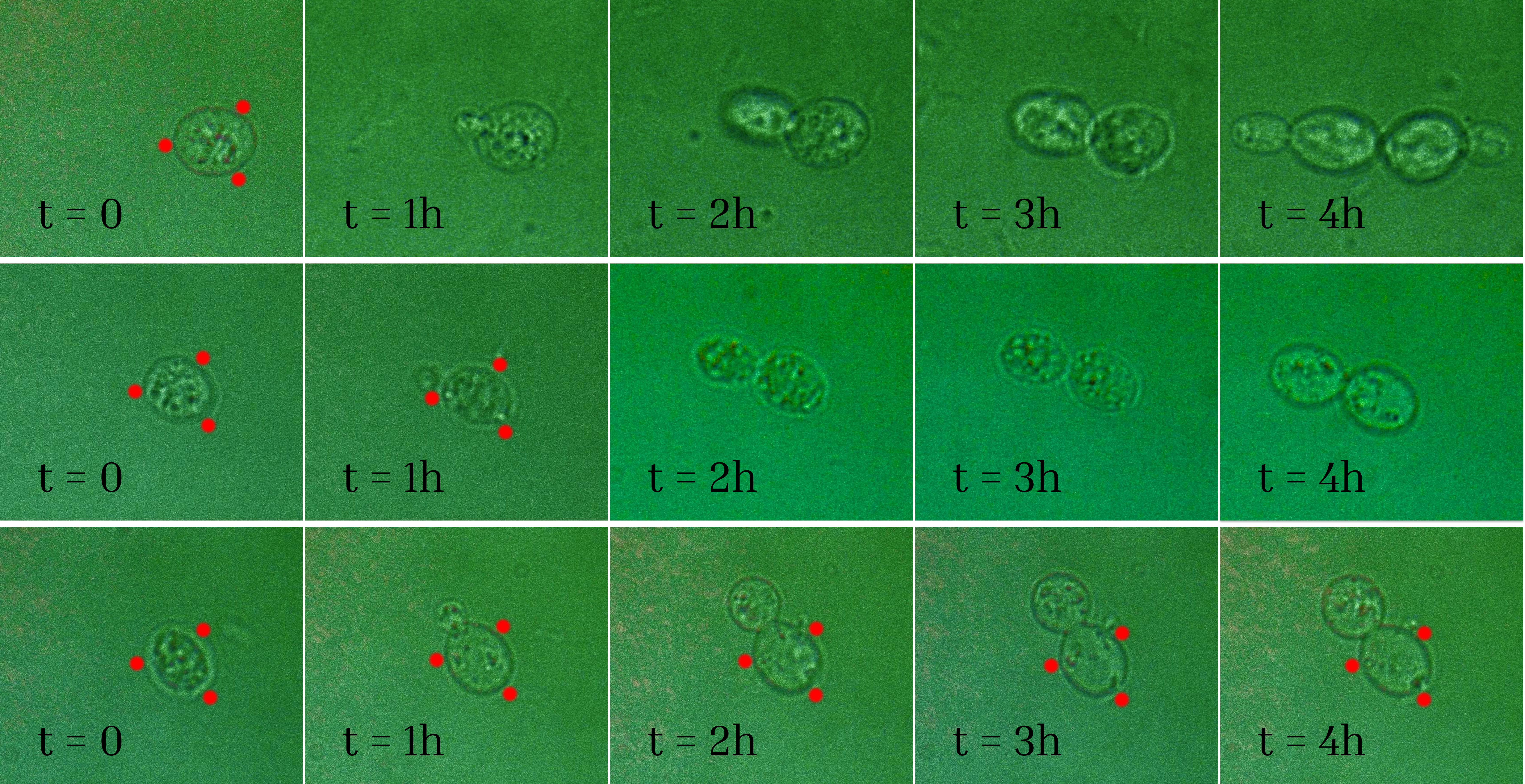}
    \caption{\textit{S. cerevisiae} reproduction in the claw dark tweezer for different trapping times prior to being released. Top row: 10 minutes; middle: 1 hour; bottom: 4 hours. The total time of observation in every experiment is 4 hours.}
    \label{fig:trapped_claw_reproduction}
\end{figure}

The area of each cell was extracted from the bounding box returned by the object-detection model described in Appendix \ref{app:yolo}. The bounding box circumscribes the cell rather than following its contour, so it overestimates the true area by a factor that depends on the cell shape. Since all the cells analyzed here have comparable shapes, and since the total area of each realization is normalized by the area of its own mother cell, this factor largely cancels out, and the normalized quantity still provides a reliable measure of the ratio between the area of a single cell and that of the resulting pair.

The results are shown in Fig.~\ref{fig:growing_graph}, each curve being the mean over five cells of approximately the same size subjected to the same protocol. Since the cells are not synchronized, individual realizations deviate from one another and we focus on the mean behavior. The vertical axis is the total area normalized by the area of the initial cell in each realization, so that it represents, approximately, the number of cells. For this strain, growth in YPG medium at \SI{30}{\celsius} is characterized by a lag phase of about 3 hours, an exponential phase between 7 and 9 hours, and a stationary phase reached at about 24 hours \cite{defalco2022luminescence,fragoso2025not}. The 4-hour window used here therefore covers the lag phase and the onset of exponential growth, which is consistent with the single duplication observed in the curves.

Cells held in the claw dark tweezer duplicate during the measurement, whereas those held in the single Gaussian tweezer remain as a single cell throughout, indicating that the claw dark tweezer is gentle enough not to arrest mitosis, in contrast with the single Gaussian tweezer, which interrupts it completely. Within the dispersion of the measurements, no significant difference is observed among the three trapping times: cells held for 4 hours reproduce as readily as cells held for 10 minutes, which indicates that the residual interaction with the Gaussian tails does not accumulate appreciably over the timescales explored here.

These results for the claw dark tweezer can also be seen directly in the images of Fig.~\ref{fig:trapped_claw_reproduction}, and in the time-lapse recording provided as Movie S3 \cite{supplemental}, which follows a single trapped cell through a complete budding event.

\section{Conclusion}

In conclusion, we have implemented a dark optical trap formed by three displaced Gaussian beams. We have confined single \textit{S. cerevisiae} cells using this claw dark tweezer operating in the regime of repulsive optical forces. Working in a water-based medium, with the proper refractive index, we confined individual yeasts and followed their position over time, comparing the results with cells held in a conventional Gaussian tweezer in the attractive regime.

Cells held in the claw dark tweezer continued to bud and duplicate over the observation window, with no appreciable dependence on the trapping time, whereas cells held in the Gaussian tweezer remained as a single cell throughout. These observations are consistent with the expectation that a trapping geometry whose equilibrium position lies in a region of negligible intensity reduces the energy absorbed by the trapped cell. The scheme can be extended to a larger number of beams, adapting the beam arrangement to other cell types and sizes, and using individual beams to deform the membrane in a controlled way are natural directions for further work. Finally, we note that viewing living cells as highly sensitive photonic sensors, our findings suggest that dark optical traps can advance quantum experiments with levitated nanoparticles by minimizing decoherence from photon recoil and thermal absorption/reemission \cite{PhysRevLett.131.163601, afridi2026controlling, illetschek2026dark}.

\section*{Acknowledgments}

We thank Bruno Melo, Daniel Tandeitnik, Oscar Schmitt Kremer, Pedro V. Paraguassú and Amanda Noronha for valuable discussions during the conception of this work. We acknowledge support from the Coordenação de Aperfeiçoamento de Pessoal de Nível Superior - Brasil (CAPES) - Finance Code 001, the Brazilian National Institute of Science and Technology in Quantum Devices (INCT-DQ) and the Brazilian National Council for Scientific and Technological Development (CNPq, Grant No. 408783/2024-9), Fundação de Amparo à Pesquisa do Estado do Rio de Janeiro (FAPERJ Scholarships No. E-26/200.251/2023, E-26/203.727/2025, E-26/210.249/2024, E-26/210.824/2025 and E-26/210.373/2026), Fundação de Amparo à Pesquisa do Estado de São Paulo (FAPESP 2021/06736-5, 2021/06823-5 e 2024/07739-6), the Serrapilheira Institute (grant No. Serra – 2211-42299) and StoneLab.

\appendix

\section{Cell viability} \label{app:cell_viability}

To evaluate the impact of the Iodixanol concentration on yeast, growth inhibition assays were conducted using the \textit{Saccharomyces cerevisiae} strain. The initial inoculum was prepared from cultures grown in liquid YPG medium (0.5\% yeast extract, 1\% peptone and 2\% glucose). The cell concentration was standardized by turbidimetry (absorbance at \SI{570}{\nano\meter}) to ensure a uniform initial optical density (0.01 turbidity units, u.t.). The exposure assay was performed in 96-well microplates, in which aliquots of the cell suspension were inoculated into YPG medium solutions containing increasing concentrations of Iodixanol (30--48\% m/v). The microplates were incubated at \SI{32}{\celsius} for 24 hours. After the incubation period, cell growth was quantified by spectrophotometry. The proliferation rate was normalized relative to the control group (cells cultured in YPG in the absence of Iodixanol), taking the control growth as 100\% viability. Statistical analysis was performed using a one-way analysis of variance (ANOVA) followed by Dunnett's multiple comparisons test to evaluate significant differences between the treated groups and the control.

\begin{figure}[htbp]
    \centering
    \includegraphics[width=0.48\textwidth]{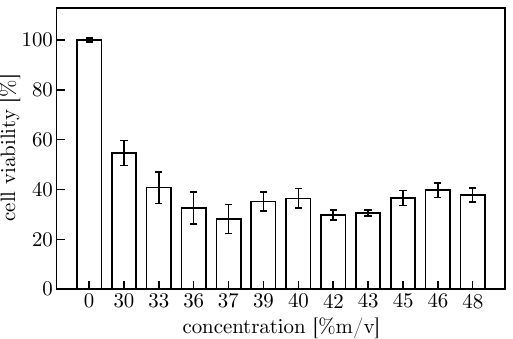}
    \caption{Viability of \textit{S. cerevisiae} after 24 hours of incubation in YPG medium containing increasing concentrations of Iodixanol, normalized to the reagent-free control. Bars show the mean and error bars the standard error of the mean.}
    \label{fig:cell_viability_rel}
\end{figure}

The growth analysis revealed that the addition of Iodixanol to the culture medium exerts an inhibitory effect on the growth of \textit{S. cerevisiae}, as shown in Fig.~\ref{fig:cell_viability_rel}. The data indicate a drop in viability as soon as the reagent is introduced. Statistical analysis revealed a significant reduction in cell viability across all tested Iodixanol concentrations when compared to the reagent-free control group ($p < 0.05$). In the solution corresponding to the working range adopted for the optical trapping experiments, which has a concentration of 40.78\% m/v of Iodixanol, the average cell growth observed was approximately 40\%. This 60\% inhibition in the proliferation rate compared to the reagent-free medium confirms that the concentration of Iodixanol imposes stress on the yeast. However, the maintenance of this 40\% growth shows that the solution is not acutely inhibitory, preserving sufficient cellular integrity to ensure a viable operational window during the laser manipulation assays.

\section{Solution composition}
\label{app:solution_composition}

For the experiments performed in this work, several solutions were tested; their compositions are listed in Table~\ref{tab:amostras}. All of them are simple dilutions of water, Optiprep (a commercial stock of 60\% Iodixanol in water), YPG (yeast extract-peptone-glucose) and a stock yeast suspension. The YPG medium was prepared following Ref.~\cite{fragoso2025not}, by dissolving peptone (1\% m/v), glucose (2\% m/v) and yeast extract (0.5\% m/v) in ultrapure water, with the pH adjusted to $6.0 \pm 0.1$; the concentrated YPG 2x used throughout this work corresponds to twice these concentrations.

The stock suspension was prepared by culturing the yeast in an aqueous YPG medium, under conditions optimal for reproduction, and a small fixed volume of this culture was transferred to each solution, so that the number of viable cells per unit volume is the same in all of them. The \textit{S. cerevisiae} strain used here, as well as its growth pattern in this medium, was isolated and characterized in previous work \cite{defalco2022luminescence,fragoso2025not}. Nine different proportions of YPG and Optiprep were tested in order to probe the repulsive response of the cells. All solutions were pre-heated to \SI{32}{\celsius} before every experiment.

\begin{table}[h]
\begin{ruledtabular}
\begin{tabular}{cccccc}
Solution & Iodixanol & Water & YPG 2x & Yeast & R.I. \\
\hline
1 & 58.25 & 38.84 & 0.00  & 2.91 & 1.426 \\
2 & 55.34 & 36.89 & 4.85  & 2.91 & 1.422 \\
3 & 52.43 & 34.95 & 9.71  & 2.91 & 1.418 \\
4 & 49.51 & 33.01 & 14.56 & 2.91 & 1.414 \\
5 & 46.60 & 31.07 & 19.42 & 2.91 & 1.410 \\
6 & 43.69 & 29.13 & 24.27 & 2.91 & 1.406 \\
\textbf{7} & \textbf{40.78} & \textbf{27.18} & \textbf{29.13} & \textbf{2.91} & \textbf{1.401} \\
8 & 37.87 & 25.24 & 33.98 & 2.91 & 1.397 \\
9 & 0.00  & 48.54 & 48.54 & 2.91 & 1.338 \\
\end{tabular}
\end{ruledtabular}
\caption{\label{tab:amostras}
Composition of solutions in volume percentage and their respective refractive indices. The YPG (yeast extract-peptone-glucose) was used in a concentrated composition, with twice the optimal concentration of it in an aqueous solution. Optiprep is added as a commercial stock of 60\% Iodixanol in water; the Iodixanol column therefore lists the iodixanol content alone, while the water introduced together with it is accounted for in the Water column. The volume fraction of Optiprep in each solution is thus the sum of the first two columns.}
\end{table}

The refractive index of each solution was computed as a volume-weighted linear combination of the indices of its components, following \cite{Boothe2017}, using $n_{\mathrm{water}} = 1.333$, $n_{\mathrm{YPG2x}} = 1.343$ and $n_{\mathrm{Optiprep}} = 1.429$, quoted at \SI{20}{\celsius}. The stock yeast suspension, being an aqueous YPG culture, was assigned the refractive index of YPG 2x in this calculation. The values in Table~\ref{tab:amostras} therefore refer to this reference temperature.

Since the medium and the cytoplasm are both predominantly aqueous, their refractive indices shift in the same direction with temperature, and the relative index $m = n_{\mathrm{cell}}/n_{\mathrm{med}}$ that determines the trapping regime is considerably less sensitive to temperature than either index separately. Together with the low absorption of water and of the cellular components at $\lambda = \SI{780}{\nano\meter}$, this justifies treating $m$ as fixed throughout the experiments.

\section{YOLO object detection model} \label{app:yolo}

Cell tracking inside the optical trap was performed by image analysis. Videos of the trapped cells were recorded for each experimental condition, and 500 frames were manually annotated using the Label Studio software to build the training dataset. A YOLOv11s (You Only Look Once) object detection model was trained in Python for 70 epochs, which was sufficient for convergence given the dataset size.

\begin{figure}[htbp]
    \centering
    \includegraphics[width=0.48\textwidth]{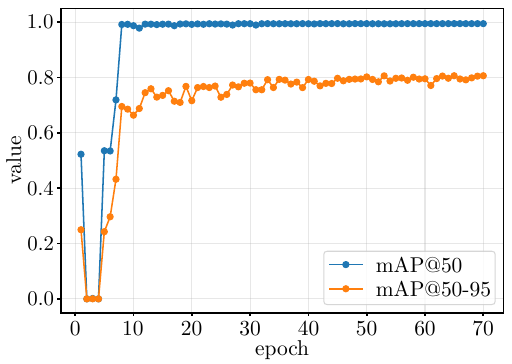}
    \caption{mAP@50 and mAP@50-95 of the YOLOv11s model across 70 training epochs.}
    \label{fig:yolo_metrics}
\end{figure}

The evolution of the two standard detection metrics used to evaluate the model is shown in Fig.~\ref{fig:yolo_metrics}. The mAP@50 (mean Average Precision at an Intersection over Union threshold of 0.5) measures how well the predicted bounding boxes match the annotated cells under a relatively lenient overlap criterion, while the mAP@50-95 averages this measure over stricter IoU thresholds ranging from 0.5 to 0.95, providing a more demanding assessment of the localization accuracy. Both metrics increase rapidly within the first 10 epochs and then stabilize, reaching values close to 1.0 for mAP@50 and approximately 0.80 for mAP@50-95, indicating that the model reliably detects and accurately localizes the cells in the trap.

\bibliography{main}

@article{Keloth2018,
  author    = {Keloth, Anusha and Anderson, Owen and Risbridger, Donald and Paterson, Lynn},
  title     = {Single Cell Isolation Using Optical Tweezers},
  journal   = {Micromachines},
  volume    = {9},
  number    = {9},
  pages     = {434},
  year      = {2018},
  month     = {August},
  doi       = {10.3390/mi9090434},
  url       = {https://doi.org/10.3390/mi9090434},
  pmid      = {30424367},
  pmcid     = {PMC6187562},
  note      = {Published: 29 August 2018}
}

@article{PhysRevApplied.14.034069,
  title = {Optical Trapping in a Dark Focus},
  author = {Melo, B. and Brand\~ao, I. and Pinheiro da, B. Silva and Rodrigues, R.B. and Khoury, A.Z. and Guerreiro, T.},
  journal = {Phys. Rev. Appl.},
  volume = {14},
  issue = {3},
  pages = {034069},
  numpages = {15},
  year = {2020},
  month = {Sep},
  publisher = {American Physical Society},
  doi = {10.1103/PhysRevApplied.14.034069},
  url = {https://link.aps.org/doi/10.1103/PhysRevApplied.14.034069}
}

@article{PhysRevLett.131.163601,
  title = {Trapping Microparticles in a Structured Dark Focus},
  author = {Almeida, F. and Sousa, I. and Kremer, O. and da Silva, B. Pinheiro and Tasca, D. S. and Khoury, A. Z. and Tempor\~ao, G. and Guerreiro, T.},
  journal = {Phys. Rev. Lett.},
  volume = {131},
  issue = {16},
  pages = {163601},
  numpages = {6},
  year = {2023},
  month = {Oct},
  publisher = {American Physical Society},
  doi = {10.1103/PhysRevLett.131.163601},
  url = {https://link.aps.org/doi/10.1103/PhysRevLett.131.163601}
}

@article{Qu2011,
  author    = {Xiaohui Qu and Jin-Der Wen and Laura Lancaster and Harry F. Noller and Carlos Bustamante and Ignacio Tinoco},
  title     = {The ribosome uses two active mechanisms to unwind messenger RNA during translation},
  journal   = {Nature},
  volume    = {475},
  pages     = {118--121},
  year      = {2011},
  month     = {July},
  doi       = {10.1038/nature10126},
  url       = {https://doi.org/10.1038/nature10126},
  note      = {Published: 06 July 2011}
}

@article{Ashkin1987,
  author    = {Ashkin, A. and Dziedzic, J. M. and Yamane, T.},
  title     = {Optical trapping and manipulation of single cells using infrared laser beams},
  journal   = {Nature},
  volume    = {330},
  pages     = {769--771},
  year      = {1987},
  month     = {December},
  doi       = {10.1038/330769a0},
  url       = {https://doi.org/10.1038/330769a0},
  note      = {Published: 31 December 1987}
}

@article{afridi2026controlling,
  title={Controlling the sign of optical forces using metaoptics},
  author={Afridi, Adeel and Melo, Bruno and Meyer, Nadine and Quidant, Romain},
  journal={Nature Communications},
  volume={17},
  number={1},
  pages={1163},
  year={2026},
  publisher={Nature Publishing Group UK London}
}

@article{hackermuller2004decoherence,
  title={Decoherence of matter waves by thermal emission of radiation},
  author={Hackerm{\"u}ller, Lucia and Hornberger, Klaus and Brezger, Bj{\"o}rn and Zeilinger, Anton and Arndt, Markus},
  journal={Nature},
  volume={427},
  number={6976},
  pages={711--714},
  year={2004},
  publisher={Nature Publishing Group UK London}
}

@article{jain2016direct,
  title={Direct measurement of photon recoil from a levitated nanoparticle},
  author={Jain, Vijay and Gieseler, Jan and Moritz, Clemens and Dellago, Christoph and Quidant, Romain and Novotny, Lukas},
  journal={Physical review letters},
  volume={116},
  number={24},
  pages={243601},
  year={2016},
  publisher={APS}
}

@article{Boothe2017,
  author    = {Boothe, Tobias and Hilbert, Lennart and Heide, Michael and Berninger, Lea and Huttner, Wieland B. and Zaburdaev, Vasily and Vastenhouw, Nadine L. and Myers, Eugene W. and Drechsel, David N. and Rink, Jochen C.},
  title     = {A tunable refractive index matching medium for live imaging cells, tissues and model organisms},
  journal   = {eLife},
  volume    = {6},
  pages     = {e27240},
  year      = {2017},
  month     = {July},
  doi       = {10.7554/eLife.27240},
  url       = {https://doi.org/10.7554/eLife.27240},
  pmid      = {28708059},
  pmcid     = {PMC5582871},
  note      = {Published: 14 July 2017}
}

@article{pilat2017effects,
  title = {Effects of Infrared Optical Trapping on Saccharomyces cerevisiae in a Microfluidic System},
  author = {Pilát, Zdeněk and Jonáš, Alexandr and Ježek, Jan and Zemánek, Pavel},
  journal = {Sensors},
  volume = {17},
  number = {11},
  pages = {2640},
  year = {2017},
  publisher = {MDPI},
  doi = {10.3390/s17112640},
  url = {https://doi.org/10.3390/s17112640}
}

@article{Ashkin1987-2,
  author  = {Ashkin, Arthur and Dziedzic, J. M.},
  title   = {Optical Trapping and Manipulation of Viruses and Bacteria},
  journal = {Science},
  volume  = {235},
  number  = {4795},
  pages   = {1517--1520},
  year    = {1987},
  doi     = {10.1126/science.3547653}
}

@article{Volpe2023,
  author = {Volpe, Giovanni and Maragò, Onofrio M. and Rubinsztein-Dunlop, Halina and Pesce, Giuseppe and Stilgoe, Alexander B. and Volpe, Giorgio and Tkachenko, Georgiy and Truong, Viet Giang and Nic Chormaic, Síle and Kalantarifard, Fatemeh and others},
  title = {Roadmap for Optical Tweezers},
  journal = {Journal of Physics: Photonics},
  volume = {5},
  number = {2},
  pages = {022501},
  year = {2023},
  doi = {10.1088/2515-7647/acb57b},
  publisher = {IOP Publishing}
}

@book{Novotny2012,
  author = {Novotny, Lukas and Hecht, Bert},
  title = {Principles of Nano-Optics},
  edition = {2},
  publisher = {Cambridge University Press},
  year = {2012}
}

@article{millen2020optomechanics,
  title={Optomechanics with levitated particles},
  author={Millen, James and Monteiro, Tania S and Pettit, Robert and Vamivakas, A Nick},
  journal={Reports on Progress in Physics},
  volume={83},
  number={2},
  pages={026401},
  year={2020},
  publisher={IOP Publishing}
}

@article{gonzalez2021levitodynamics,
  title={Levitodynamics: Levitation and control of microscopic objects in vacuum},
  author={Gonzalez-Ballestero, Carlos and Aspelmeyer, Markus and Novotny, Lukas and Quidant, Romain and Romero-Isart, Oriol},
  journal={Science},
  volume={374},
  number={6564},
  pages={eabg3027},
  year={2021},
  publisher={American Association for the Advancement of Science}
}

@article{gieseler2014dynamic,
  title={Dynamic relaxation of a levitated nanoparticle from a non-equilibrium steady state},
  author={Gieseler, Jan and Quidant, Romain and Dellago, Christoph and Novotny, Lukas},
  journal={Nature nanotechnology},
  volume={9},
  number={5},
  pages={358--364},
  year={2014},
  publisher={Nature Publishing Group UK London}
}

@article{trepagnier2004experimental,
  title={Experimental test of Hatano and Sasa's nonequilibrium steady-state equality},
  author={Trepagnier, EH and Jarzynski, Christopher and Ritort, Felix and Crooks, Gavin E and Bustamante, CJ and Liphardt, J},
  journal={Proceedings of the National Academy of Sciences},
  volume={101},
  number={42},
  pages={15038--15041},
  year={2004},
  publisher={National Academy of Sciences}
}

@article{rieser2022tunable,
  title={Tunable light-induced dipole-dipole interaction between optically levitated nanoparticles},
  author={Rieser, Jakob and Ciampini, Mario A and Rudolph, Henning and Kiesel, Nikolai and Hornberger, Klaus and Stickler, Benjamin A and Aspelmeyer, Markus and Deli{\'c}, Uro{\v{s}}},
  journal={Science},
  volume={377},
  number={6609},
  pages={987--990},
  year={2022},
  publisher={American Association for the Advancement of Science}
}

@article{penny2023sympathetic,
  title={Sympathetic cooling and squeezing of two colevitated nanoparticles},
  author={Penny, TW and Pontin, A and Barker, PF},
  journal={Physical Review Research},
  volume={5},
  number={1},
  pages={013070},
  year={2023},
  publisher={APS}
}

@article{illetschek2026dark,
  title={Dark Optical Trapping of Resonant Transition-Metal Dichalcogenide Particles},
  author={Illetschek, Patrick and Fedorovich, Gleb and Seredin, Albert and Tselikov, Gleb and Volkov, Valentin S and Kiesel, Nikolai and Aspelmeyer, Markus and Petrov, Mihail and Zasedatelev, Anton V},
  journal={arXiv preprint arXiv:2607.09896},
  year={2026}
}

@article{mohanty2005dynamics,
  title={Dynamics of Interaction of RBC with optical tweezers},
  author={Mohanty, Samarendra K and Mohanty, Khyati S and Gupta, Pradeep K},
  journal={Optics express},
  volume={13},
  number={12},
  pages={4745--4751},
  year={2005},
  publisher={Optical Society of America}
}

@article{wilson2025free,
  title={On the free energy of protein folding in optical tweezers experiments},
  author={Wilson, Christian AM and Corr{\^e}a, Camila G},
  journal={Biophysical Reviews},
  volume={17},
  number={2},
  pages={231--245},
  year={2025},
  publisher={Springer}
}

@article{collin2005verification,
  title={Verification of the Crooks fluctuation theorem and recovery of RNA folding free energies},
  author={Collin, Delphine and Ritort, Felix and Jarzynski, Christopher and Smith, Steven B and Tinoco Jr, Ignacio and Bustamante, Carlos},
  journal={Nature},
  volume={437},
  number={7056},
  pages={231--234},
  year={2005},
  publisher={Nature Publishing Group UK London}
}

@article{kremer2024perturbative,
  title={Perturbative nonlinear feedback forces for optical levitation experiments},
  author={Kremer, Oscar and Tandeitnik, Daniel and Mufato, Rafael and Califrer, Igor and Calderoni, Breno and Calliari, Felipe and Melo, Bruno and Tempor{\~a}o, Guilherme and Guerreiro, Thiago},
  journal={Physical Review A},
  volume={109},
  number={2},
  pages={023521},
  year={2024},
  publisher={APS}
}

@article{pikovski2012probing,
  title={Probing Planck-scale physics with quantum optics},
  author={Pikovski, Igor and Vanner, Michael R and Aspelmeyer, Markus and Kim, MS and Brukner, {\v{C}}aslav},
  journal={Nature Physics},
  volume={8},
  number={5},
  pages={393--397},
  year={2012},
  publisher={Nature Publishing Group UK London}
}

@article{weiss2021large,
  title={Large quantum delocalization of a levitated nanoparticle using optimal control: Applications for force sensing and entangling via weak forces},
  author={Weiss, T and Roda-Llordes, M and Torrontegui, E and Aspelmeyer, M and Romero-Isart, O},
  journal={Physical Review Letters},
  volume={127},
  number={2},
  pages={023601},
  year={2021},
  publisher={APS}
}

@article{dania2025high,
  title={High-purity quantum optomechanics at room temperature},
  author={Dania, Lorenzo and Kremer, Oscar Schmitt and Piotrowski, Johannes and Candoli, Davide and Vijayan, Jayadev and Romero-Isart, Oriol and Gonzalez-Ballestero, Carlos and Novotny, Lukas and Frimmer, Martin},
  journal={Nature Physics},
  volume={21},
  number={10},
  pages={1603--1608},
  year={2025},
  publisher={Nature Publishing Group UK London}
}

@article{ashkin1986observation,
  title={Observation of a single-beam gradient force optical trap for dielectric particles},
  author={Ashkin, Arthur and Dziedzic, James M and Bjorkholm, John E and Chu, Steven},
  journal={Optics letters},
  volume={11},
  number={5},
  pages={288--290},
  year={1986},
  publisher={Optical Society of America}
}

@article{neuman2004optical,
  title={Optical trapping},
  author={Neuman, Keir C and Block, Steven M},
  journal={Review of scientific instruments},
  volume={75},
  number={9},
  pages={2787--2809},
  year={2004},
  publisher={American Institute of Physics}
}

@article{bustamante2021optical,
  title={Optical tweezers in single-molecule biophysics},
  author={Bustamante, Carlos J and Chemla, Yann R and Liu, Shixin and Wang, Michelle D},
  journal={Nature Reviews Methods Primers},
  volume={1},
  number={1},
  pages={25},
  year={2021},
  publisher={Nature Publishing Group UK London}
}

@article{bianco2023label,
  title={Label-free intracellular multi-specificity in yeast cells by phase-contrast tomographic flow cytometry},
  author={Bianco, Vittorio and D'Agostino, Massimo and Pirone, Daniele and Giugliano, Giusy and Mosca, Nicola and Di Summa, Maria and Scerra, Gianluca and Memmolo, Pasquale and Miccio, Lisa and Russo, Tommaso and others},
  journal={Small Methods},
  volume={7},
  number={11},
  pages={2300447},
  year={2023},
  publisher={Wiley Online Library}
}

@article{liu2016cell,
  title={Cell refractive index for cell biology and disease diagnosis: past, present and future},
  author={Liu, Patricia Yang and Chin, Lip Ket and Ser, Wee and Chen, HF and Hsieh, C-M and Lee, C-H and Sung, K-B and Ayi, TC and Yap, PH and Liedberg, Bo and others},
  journal={Lab on a Chip},
  volume={16},
  number={4},
  pages={634--644},
  year={2016},
  publisher={The Royal Society of Chemistry}
}

@article{rappaz2005measurement,
  title={Measurement of the integral refractive index and dynamic cell morphometry of living cells with digital holographic microscopy},
  author={Rappaz, Benjamin and Marquet, Pierre and Cuche, Etienne and Emery, Yves and Depeursinge, Christian and Magistretti, Pierre J},
  journal={Optics express},
  volume={13},
  number={23},
  pages={9361--9373},
  year={2005},
  publisher={Optical Society of America}
}

@article{blazquez2019optical,
  title={Optical tweezers: Phototoxicity and thermal stress in cells and biomolecules},
  author={Bl{\'a}zquez-Castro, Alfonso},
  journal={Micromachines},
  volume={10},
  number={8},
  pages={507},
  year={2019},
  publisher={MDPI}
}

@article{zhang2008optical,
  title={Optical tweezers for single cells},
  author={Zhang, Hu and Liu, Kuo-Kang},
  journal={Journal of the Royal Society interface},
  volume={5},
  number={24},
  pages={671},
  year={2008}
}

@article{rademacher2026roto,
  title={Roto-translational levitated optomechanics},
  author={Rademacher, M and Pontin, A and Gosling, JMH and Barker, PF and Toro{\v{s}}, M},
  journal={Physics Reports},
  volume={1187},
  pages={1--66},
  year={2026},
  publisher={Elsevier}
}

@article{botstein2011yeast,
  title={Yeast: an experimental organism for 21st century biology},
  author={Botstein, David and Fink, Gerald R},
  journal={Genetics},
  volume={189},
  number={3},
  pages={695--704},
  year={2011},
  publisher={Oxford University Press}
}

@article{herskowitz1988life,
  title={Life cycle of the budding yeast Saccharomyces cerevisiae},
  author={Herskowitz, Ira},
  journal={Microbiological reviews},
  volume={52},
  number={4},
  pages={536--553},
  year={1988}
}

@article{fragoso2025not,
  title   = {Not that Innocent: Chemical and Toxicological Evaluation of Glycerin and Propylene Glycol Used in Vape Liquid Production},
  author  = {Fragoso, Carlos L. R. and De Falco, Anna and Santa-Helena, Eduarda and Espinosa, Guilherme V. and Gioda, Carolina R. and Gioda, Adriana},
  journal = {Journal of the Brazilian Chemical Society},
  volume  = {36},
  number  = {3},
  pages   = {e-20240222},
  year    = {2025},
  doi     = {10.21577/0103-5053.20240222},
  url     = {https://doi.org/10.21577/0103-5053.20240222},
  publisher = {Sociedade Brasileira de Química}
}

@article{defalco2022luminescence,
  title   = {Luminescence imaging and toxicity assessment of graphene quantum dots using in vitro models},
  author  = {De Falco, Anna and Santa-Helena, Eduarda and Toloza, Carlos A. T. and Almeida, Joseany M. S. and Larrude, Dunieskys G. and Meirelles, Fatima Ventura Pereira and Gioda, Carolina Rosa and Aucelio, Ricardo Q. and Gioda, Adriana},
  journal = {Fullerenes, Nanotubes and Carbon Nanostructures},
  volume  = {30},
  number  = {6},
  pages   = {657--666},
  year    = {2022},
  doi     = {10.1080/1536383X.2021.1995367},
  url     = {https://doi.org/10.1080/1536383X.2021.1995367},
  publisher = {Taylor \& Francis}
}

@misc{supplemental,
  note = {See Supplemental Material at [URL will be inserted by publisher]
  for Movie S1, showing the repulsive interaction between a single Gaussian
  beam and a yeast cell; Movie S2, showing the confinement and subsequent
  release of a single cell in the claw dark tweezer; and Movie S3, a
  time-lapse of the reproduction of a trapped cell over four hours.}
}
\end{document}